# Graded resonator arrays for spatial frequency separation and amplification of water waves

Luke G. Bennetts[1] Malte A. Peter[2,3] and Richard V. Craster[4]

[1]School of Mathematical Sciences, University of Adelaide, Adelaide, SA 5005, Australia

[2]Institute of Mathematics, University of Augsburg, 86135 Augsburg, Germany

[3]Augsburg Centre for Innovative Technologies, University of Augsburg, 86135 Augsburg, Germany

[4]Department of Mathematics, Imperial College London, South Kensington Campus, London, SW7 2AZ, UK

A structure capable of substantially amplifying water waves over a broad range of frequencies at selected locations is proposed. The structure consists of a small number of C-shaped cylinders arranged in a line array, with the cylinder properties graded along the array. Using linear potential-flow theory, it is shown that the energy carried by a plane incident wave is amplified within specified cylinders, for wavelengths comparable to the array length, and for a range of incident directions. Transfer matrix analysis is used to attribute the large amplifications to excitation of Rayleigh–Bloch waves and gradual slowing down of their group velocity along the array.

## 1. Introduction

Arrays of fixed or floating bodies are common in contemporary water-wave problems, including wave-energy harvesting Scruggs and Jacob (2009), offshore wind farms Prez-Collazo *et al.* (2015), coastal protection Martinelli *et al.* (2008) and supports for bridges or other offshore structures Faltinsen (1990). An overarching challenge is to design arrays that control the spatial distribution of wave energy, and, particularly in the case of wave-energy harvesting, to amplify the wave energy of target frequencies at the locations of the array elements (in this context the wave-energy converters; e.g. Falnes 1980; Mavrakos and McIver 1997; Göteman 2017).

An apparently disconnected subject is that of wave propagation through metamaterials, where the term metamaterial specifically refers to use of sub-wavelength resonator arrays, as opposed to Bragg scattering in periodic crystal lattices. Metamaterials originate in optics/electromagnetism Pendry *et al.* (2006), and have been used to realise remarkable behaviours, such as cloaking Schurig *et al.* (2006) and super-resolution lenses Pendry (2000), among many others. These successes have motivated development of metamaterials in other areas of physics Wegener (2013), most notably in acoustics (e.g. Liu *et al.* 2000), and elasticity/seismology (e.g. Brûlé *et al.* 2014).

Metamaterials have found relatively few applications in the water-wave context, with some notable exceptions. Most relevant to the present study, Hu *et al.* (2011) theoretically predict negative effective gravity within a doubly-periodic array of resonators in the form of thin-walled, hollow, bottom-mounted, vertical cylinders with narrow slits, analogous to split-ring resonators familiar in optics or Helmholtz resonators in acoustics (e.g. Schurig *et al.* 2006). The negative gravity prohibits propagation of low-frequency water waves, which Hu *et al.* (2013) demonstrate experimentally using a line of C-shaped cylinders (i.e. single slits) in a wave flume. In a similar study, Dupont *et al.* (2017) use numerical simulations and laboratory experiments to show that a doubly-periodic array of C-shaped cylinders prohibits low-frequency wave propagation.

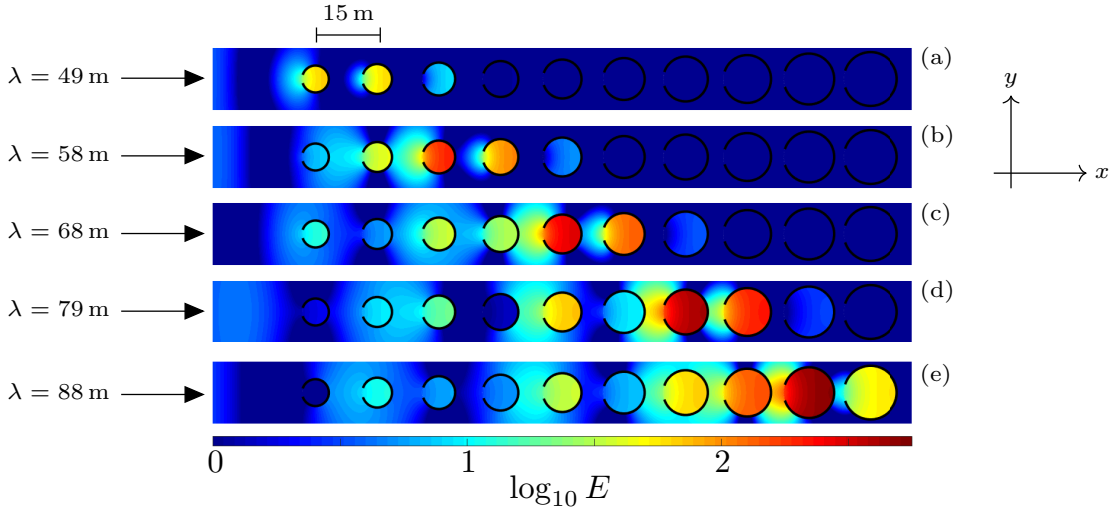

FIGURE 1. Logarithm of normalised energy along an array of C-shaped cylinders for head-on incident waves (propagating in the positive $x$-direction) with wavelength (a) $\lambda = 49\,\text{m}$, (b) $58\,\text{m}$, (c) $68\,\text{m}$, (d) $79\,\text{m}$ and (e) $88\,\text{m}$, corresponding to resonant wavelengths in cylinder 2, 4, 6, 8 and 10, respectively (ordered left to right), when in isolation.

Graded arrays, also known as chirped arrays, are an active area in metamaterials research at present (and also in photonic/phononic-crystal research). The general concept is that for a graded structure, surface, or waveguide, different frequencies can be trapped, and hence amplified, at different spatial locations, thus creating a device that traps a broadband signal. This is often referred to as rainbow trapping, due to the spatial frequency separation Tsakmakidis *et al.* (2007). In the case of a metamaterial, sub-wavelength regimes can be accessed and the device spatially separates the differing frequency components by grading the sub-wavelength resonators. The locations at which different wavelengths are isolated are then controlled and predicted using knowledge of the individual resonances and their interactions, with applications to, e.g., broadband absorption of light Narimanov and Kildishev (2009) and sound Jimenez *et al.* (2016).

In this study, a method is illustrated to amplify water wave energy along a structure, at locations selected according to wave frequency, using resonant structural elements and grading the element properties along the structure. In comparison with most cognate studies in optics/acoustics/elasticity, and in keeping with practical design considerations in the field of water waves, the structure contains a modest number of elements only, consisting of a single line array of elements, and with the array length comparable to the target wavelengths. The elements are bottom-mounted C-shaped cylinders, similar to those used by Hu *et al.* (2013) and Dupont *et al.* (2017), although, crucially, the cylinder properties are graded in the present study.

As a motivating example, Fig. 1 shows the depth-integrated wave energy, $E$, along a line array of ten C-shaped cylinders in a water domain of infinite horizontal extent $(x, y \in \mathbb{R})$, produced by plane incident waves travelling in the direction of the line (the $x$-direction, i.e. head-on incidence). Energy distributions are shown for five different wavelengths, $\lambda$, where the cylinder radius gradually increases from $3.25\,\text{m}$ to $6.5\,\text{m}$, and the incident energy is normalised to unity. To put the array dimensions in context, adjacent cylinder centres are $15\,\text{m}$ apart, which is less than three times smaller than the shortest wavelength considered, and means the overall length of the array is 142.75 m, which is less than three times greater than the shortest wavelength and less than two times the longest wavelength.

The five incident wavelengths are chosen to be close to the longest-resonant wavelengths for cylinders 2, 4, 6, 8 and 10 (ordered from left to right). Large energy amplifications are evident in these cylinders for the corresponding wavelengths, but are often overshadowed by even larger amplifications in the cylinders immediately preceding them (with respect to the incident wave direction). The maximum amplifications increase with increasing wavelength, from $E \approx 67.2$ in cylinder 1 for $\lambda = 49$ m, to $E \approx 524$ in cylinder 9 for $\lambda = 88$ m, which is over 11 times greater than the amplification for the cylinder in isolation. These significant amplifications have been attained without invoking optimisation strategies or parameter tuning.

## 2. Preliminaries

Consider a water domain of infinite horizontal extent and finite depth $h$, bounded below by a flat bed and above by a free surface, and containing a line array of $M$ infinitely thin, vertical, C-shaped cylinders that extend throughout the water column. A Cartesian coordinate system $\mathbf{X} = (x, y, z)$ defines locations in the water domain, where $\mathbf{x} = (x, y)$ is the horizontal coordinate and $z$ is the vertical coordinate. The vertical coordinate points directly upwards, with its origin, $z = 0$, coinciding with the equilibrium free surface, and $z = -h$ denoting the bed.

Under the usual assumptions of linear water-wave theory (incompressible, inviscid and irrotational fluid), the water velocity field is defined as $\mathbf{u}(\mathbf{X}, t) = \nabla \operatorname{Re}\{(g/\mathrm{i}\omega)\phi(\mathbf{X})\exp(-\mathrm{i}\omega t)\}$, where $\phi$ is a reduced complex-valued velocity potential, $\omega \in \mathbb{R}$ is a prescribed angular frequency, and $g \approx 9.81\,\mathrm{m\,s^{-2}}$ is the constant of gravitational acceleration. The velocity potential satisfies Laplace's equation throughout the linearised water domain, i.e.

$$\nabla^2 \phi = 0 \quad \text{for} \quad \mathbf{X} \in \mathcal{D} \times (-h, 0), \tag{2.1a}$$

where $\mathcal{D} = \mathbb{R}^2 \setminus \overline{\mathcal{C}}$, and $\mathcal{C}$ is the union of the $M$ C-shaped cylinder horizontal cross-sections. It also satisfies the impermeable bed and free surface conditions,

$$\frac{\partial \phi}{\partial z} = 0 \quad \text{for} \quad \mathbf{x} \in \mathcal{D},\ z = -h, \quad \text{and} \quad \frac{\partial \phi}{\partial z} = \frac{\omega^2}{g}\phi \quad \text{for} \quad \mathbf{x} \in \mathcal{D},\ z = 0, \tag{2.1b}$$

respectively, and the Sommerfeld radiation condition in the far field $|\mathbf{x}| \to \infty$.

Without loss of generality, the array is assumed to lie along the $x$-axis. Let the cylinders be indexed $m = 1, \ldots, M$ from left to right, and the domain occupied by the $m$th cylinder be $\mathbf{X} \in \mathcal{C}_m \times (-h, 0)$, where

$$\mathcal{C}_m = \{\mathbf{x} : (x - x_m)^2 + y^2 = a_m^2, \arctan(y/(x - x_m)) < \pi - \varphi\}. \tag{2.2}$$

Here, $x_m = (m - 1)W$ is the cylinder centre location along the $x$-axis, $a_m$ is its radius, and $\varphi$ is the half-angle of its opening (identical for all cylinders), and the openings are at the left-hand end of the cylinders and symmetric about the $x$-axis, as shown in Fig. 1. The velocity potential satisfies a no-normal-flow condition at the cylinder surfaces, i.e.

$$\frac{\partial \phi}{\partial n} = 0 \quad \text{for} \quad \mathbf{X} \in \mathcal{C}_m \times (-h, 0) \quad (m = 1, \ldots, M), \tag{2.3}$$

where $\partial/\partial n \equiv \mathbf{n} \cdot \nabla$ and $\mathbf{n}$ is the normal vector to the cylinder surfaces, together with a condition ensuring the correct singularity at the tips of the C-shape.

Motions are forced by a unit-amplitude plane incident wave, with velocity potential

$$\phi_{\text{inc}}(\mathbf{X}) = \exp\{i\,k\,(x\cos\psi + y\sin\psi)\}\,\frac{\cosh\{k\,(z+h)\}}{\cosh(k\,h)}, \qquad (2.4)$$

where $\psi$ is the incident wave direction with respect to the positive $x$-axis ($\psi = 0$ for head-on incidence, as in Fig. 1), and the wavenumber $k = 2\pi/\lambda \in \mathbb{R}_+$ satisfies the dispersion relation $k\tanh(k\,h) = \omega^2/g$, and is therefore used as a proxy for frequency. The full wave field is $\phi = \phi_{\text{inc}} + \phi_{\text{sca}}$, where $\phi_{\text{sca}}$ is the scattered wave field.

The standard solution method for this type of problem is to calculate so-called diffraction transfer matrices, which encode the scattering by the individual cylinders, and use Graf's addition formula to calculate wave interactions between cylinders Kagemoto and Yue (1986); Peter and Meylan (2004). Instead, the transfer matrix method of Bennetts *et al.* (2017) is employed, as it provides insights into the mechanisms underlying the large amplifications shown in Fig. 1. Specifically, the $x$-axis containing the cylinders is divided into $M$ contiguous subintervals $(x_{m-1/2}, x_{m+1/2})$ for $m = 1, \ldots, M$, where $x_{m\pm 1/2} = x_m \pm W/2$, so that the $m$th subinterval contains cylinder $m$ only. The wave field in the $m$th subinterval is expressed in the directional-spectrum form

$$\phi = \int_{\Gamma_\pm} A_m^\pm(\chi)\,e^{i\,k\,\{(x-x_m)\cos\chi + y\sin\chi\}}\,d\chi + \int_{\Gamma_\mp} B_m^\pm(\chi)\,e^{i\,k\,\{(x-x_m)\cos\chi + y\sin\chi\}}\,d\chi \quad (2.5)$$

for $0 \leq \pm(x - x_m) \leq W/2$ and $\mathbf{x} \notin \overline{\Omega}_m$, where $\Omega_m = \{\mathbf{x} : (x - x_m)^2 + y^2 < a_m\}$, $\Gamma_- = \{-\pi/2 + i\gamma : \gamma \in \mathbb{R}_+\} \cup \{\gamma \in \mathbb{R} : -\pi/2 \leq \gamma \leq \pi/2\} \cup \{\pi/2 - i\gamma : \gamma \in \mathbb{R}_+\}$ and $\Gamma_+ = \Gamma_- + \pi$. Transfer matrices $\mathbf{P}_m$ are calculated, which map discretised versions of the amplitude functions $A_m^\pm$ and $B_m^\pm$ from the left ($-$) to the right ($+$) of the subintervals. The amplitudes, and hence solution, are then found recursively, moving from the leftmost subinterval to rightmost, and applying radiation conditions at the ends.

## 3. Rayleigh–Bloch waves, dispersion curves and quasi-bandgaps

Fig. 2 shows results that provide insights into the large amplifications observed in Fig. 1, for which the array is defined by $M = 10$, $W = 15\,\text{m}$, $\varphi = 0.1\,\pi$, and $a_m = a_1\,(m + M - 2)/(M - 1)$ where $a_1 = 3.25\,\text{m}$. The largest amplification case from Fig. 1, with $\lambda = 88\,\text{m}$, is chosen as an example, and Fig. 2a is a magnified version of Fig. 1e.

As shown by Thompson *et al.* (2008) and others for uniform line arrays of regular cylinders, the scattered wave field along the array is dominated by so-called Rayleigh–Bloch waves, which propagate in both directions along the array and decay exponentially in the transverse direction (i.e. $y$-direction) away from it. As explained below, line arrays of C-shaped cylinders also support Rayleigh–Bloch waves, and the radius grading causes their properties to evolve along the array. Fig. 2b is similar to Fig 2a, but with the scattered wave field approximated by the Rayleigh–Bloch wave components only. The approximation is based on projecting the scattered wave field in each subinterval onto the eigenfunctions defined by the corresponding transfer matrix, and retaining only the eigenfunctions associated to the Rayleigh–Bloch waves. Notwithstanding the small discontinuities, which are an inevitable consequence of the approximation method, the approximation is highly accurate, particularly with respect to the large amplifications, thus confirming the amplifications are due to excitation of Rayleigh–Bloch waves.

Figs. 2c–g show eigenvalue spectra of the transfer matrices, $\mu \in \text{eig}\{\mathbf{P}_m\}$, for $m = 2, 5, 7, 9$ and $10$, respectively, in the complex plane. The closely spaced black dots along an arc of the unit circle represent the continuous spectrum, corresponding to plane wave forcing, i.e. $\mu = \exp\{i\,k\,W\cos\chi\}$ for $\chi \in (-\pi, \pi)$. (Eigenvalues in the continuous spectrum

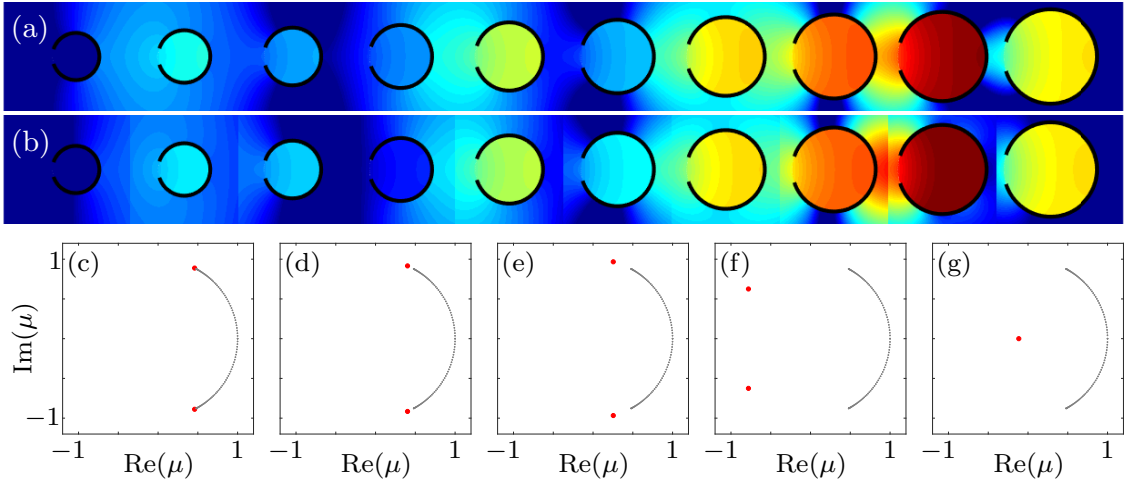

FIGURE 2. (a) Log of normalised energy as in Fig. 1e. (b) Corresponding energy with scattered wave field approximated by the Rayleigh–Bloch-wave components only. (c–g) Eigenvalue spectra of transfer matrices operating over (c) cylinder $m = 2$, (d) 5, (e) 7, (f) 9 and (g) 10, where the closely spaced black dots define the discrete representation of the continuous spectra and the red dots define the discrete spectra corresponding to Rayleigh–Bloch waves.

corresponding to evanescent wave forcing are omitted from the plots for clarity.) The two red dots represent the discrete spectrum, which correspond to Rayleigh–Bloch waves, i.e. $\mu = \exp\{\pm i\,\beta\,W\}$, where $\beta$ is the Rayleigh–Bloch wavenumber, with the eigenvalue in the upper-half complex plane corresponding to a rightward-propagating Rayleigh–Bloch wave $(+)$ and the eigenvalue in the lower-half plane corresponding to a leftward-propagating Rayleigh–Bloch wave $(-)$.

Fig. 2c shows the spectrum for the transfer matrix which maps over cylinder $m = 2$, for which the Rayleigh–Bloch eigenvalues are fractionally displaced along the unit circle from the ends of the continuous spectrum. Figs. 2d–f show that the spectra of the transfer matrices farther along the array have identical continuous spectra, but that the Rayleigh–Bloch eigenvalues rapidly move along the unit circle away from the continuous spectrum. As the Rayleigh–Bloch eigenvalues approach $-1$ from above $(+)$ and below $(-)$, the group velocity of the Rayleigh–Bloch waves reduces, and when the eigenvalues meet at $-1$ the Rayleigh–Bloch waves become standing waves known as a Neumann trapped mode Evans and Porter (1999). Fig. 2g shows the spectrum for the final cylinder along the array ($m = 10$), for which the Rayleigh–Bloch eigenvalues have jumped onto the negative branch of the real line, meaning $\beta \in \pi + i\,\mathbb{R}_+$ and the Rayleigh–Bloch waves no longer propagate. (The eigenvalue associated to $-\beta$ is beyond the axes limits for $m = 10$.)

Therefore, the cylinder-radius grading causes the rightward-propagating Rayleigh–Bloch wave excited at the leading end of the array ($m = 1$) to slow down progressively along the array, until it reaches a turning point, at which the group velocity is zero. In the language of waveguide modes, at this point, the wave is cut-off and ceases to propagate. The energy carried by the Rayleigh–Bloch wave accumulates at the turning point, thereby generating large amplifications. This behaviour is broadly analogous to the effect of gradually decreasing the width of an acoustic or ocean waveguide, so that a waveguide mode is eventually cut-off with its energy reflected from the turning point, with propagation in a wedge being the archetypal example (Arnold and Felsen 1983).

The location of the Rayleigh–Bloch eigenvalues in the complex plane depends on the frequency of motion, i.e. incident wavelength. In particular, as frequency increases the eigenvalues depart the unit circle for smaller cylinder radii, meaning the Rayleigh–Bloch waves propagate shorter distances along the array. Thus, as shown in Fig. 1, energy accumulates closer to the leading end of the array as frequency increases, i.e. incident wavelength decreases, so that the array spatially separates frequencies/wavelengths.

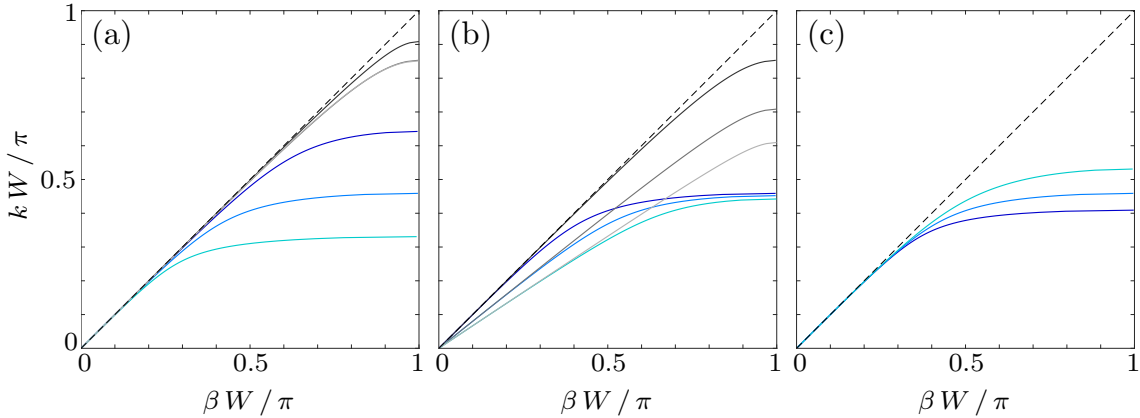

FIGURE 3. Rayleigh–Bloch dispersion curves for C-shaped cylinders (solid blue curves) and regular cylinders (solid grey), with the light-line (broken black) given for reference. (a) Varying cylinder radius: $a = 3.25$ m (i.e. cylinder $m = 1$; –/–), $a = 4.7$ m ($m = 5$; –/–) and $a = 6.5$ m ($m = 10$; –/–). (b) Varying cylinder spacing for $a = 4.7$ m: $W = 15$ m (–/–), $W = 18.75$ m (–/–) and $a = 22.5$ m (–/–). (c) Varying opening half-angle: $\varphi = \pi/20$ (–), $\varphi = \pi/10$ (–) and $\varphi = \pi/5$ (–).

Fig. 3 shows Rayleigh–Bloch dispersion curves for C-shaped cylinders in periodic line arrays. These curves are valuable for interpreting behaviours along graded arrays as, assuming the array grading is sufficiently gradual that waves behave locally as if they are in a uniform array, one can then use the dispersion curves to estimate the turning points for different frequencies. The dispersion curves are shown in the first irreducible Brillouin zone $\beta W \in [0, \pi]$, and curves for regular cylinders are shown for comparison where appropriate, along with the dispersion line (so-called light-line) for the bulk media, $k = \beta$. Fig. 3a shows dispersion curves corresponding to the array used in Figs. 1–2. Each C-shaped cylinder curve follows the light-line for low frequencies $k$/small wavenumbers $\beta$, but rapidly departs the light-line as the frequency/wavenumber increases, and cuts off ($\beta W = \pi$) at a relatively low frequency, just below the resonant frequency for the isolated cylinder, meaning the cut-off can be tuned by altering the resonance.

The frequency interval occupied by the dispersion curve is conventionally known as a passband in periodic media. In this instance, frequencies in the passband support propagating Rayleigh–Bloch waves. For frequencies above the passband, Rayleigh–Bloch waves do not propagate along the array, which is analogous to a so-called bandgap in periodic media, but is here referred to as a quasi-bandgap due to propagating modes in the continuous spectrum being supported by the array for these frequencies (as shown in Fig. 2g). As the radius of the cylinders increases, the resonant frequency decreases, thus pushing the dispersion curves down, and narrowing the passbands. Therefore, a range of frequencies exist in the passband for the small cylinders at the leading end of the array that transition into quasi-bandgaps as the radius increases along the array. Rayleigh–Bloch waves at these frequencies meet a turning point within the array, with the associated energy trapped within the array. Doubling the radius from $a = 3.25$ m to 6.5 m almost halves the passband width (reduction factor $\approx 0.51$), indicating that radius grading is an effective way to capture a wide range of frequencies.

For the regular cylinders, the dispersion curves follow the light-line for a wide range of frequencies/wavenumbers in the Brillouin zone, before departing the light-line and cutting-off just below $kW = \pi$. Therefore, only relatively high-frequency/short wavelengths fail to excite propagating Rayleigh–Bloch waves along the array, and thus can be captured. Moreover, doubling the radius of the regular cylinders from $a = 3.25$ m to $6.5$ m reduces the passband width by a factor of $\approx 0.94$ only, and increasing the cylinder from $4.7$ m to $6.5$ m reduces the width by less than 1% (curves are virtually indistinguishable). It follows that grading the radius of regular cylinders captures only a narrow range of (high) frequencies and is less effective than using the C-shaped resonators.

Fig. 3b shows the effect of changing cylinder spacing, $W$, on the dispersion curves, for a cylinder radius $a = 4.7$ m, i.e. cylinder $m = 5$ in Figs. 1–2, noting that different radii produce cognate results. For the C-shaped cylinders, the passband width is insensitive to changes in the array spacing (increasing from $W = 15$ m to $22.5$ m reduces the width by $\approx 4\%$ only), as the upper limit of the passband is controlled by the resonant frequency for an isolated cylinder, which is independent of the spacing. Therefore, grading the spacing of identical C-shaped cylinders is an ineffective approach to capture a wide range of frequencies. In contrast, quasi-bandgaps for regular cylinders are generated by Bragg scattering mechanisms, so that passband widths can be controlled by the array spacing, noting that this approach is used in chirped sonic arrays (Romero-García *et al.* 2013; Cebrecos *et al.* 2014). Increasing the spacing from $W = 15$ m to $22.5$ m reduces the passband width by a factor of $\approx 0.71$, providing a wide frequency capture range, albeit for relatively high frequencies.

Fig. 3c shows the effect of changing the cylinder opening angle on the dispersion curves, which is relevant for the C-shaped cylinders only. Decreasing the half-angle from $\varphi = \pi/5$ to $\pi/20$ decreases the passband width by an appreciable factor of $\approx 0.77$, although the reduction is small in comparison to the proportional changes in passband width given by varying the cylinder radius. Therefore, grading the opening angle along the array provides an alternative method to generate turning points along the array, and could potentially be combined with radius grading to enhance the frequency capture width (this is not pursued further here).

## 4. Amplification spectra

Fig. 4 quantifies the overall energy amplification produced by the graded array over ranges of incident wavelengths and directions, using metrics analogous to the Q-factor familiar in assessing energy gains (or losses) given by arrays of interacting wave-energy converters Falnes (1980). Fig. 4a shows $\mathcal{Q}_{\text{arr}} = \mathcal{E}/\mathcal{E}_{\text{inc}}$ on a logarithmic scale, where

$$\mathcal{E} = \sum_{m=1}^{M} \iint_{\Omega_m} |\phi|^2 \, \mathrm{d}\mathbf{x} \quad \text{and} \quad \mathcal{E}_{\text{inc}} = \sum_{m=1}^{M} \iint_{\Omega_m} |\phi_{\text{inc}}|^2 \, \mathrm{d}\mathbf{x} = \sum_{m=1}^{M} \pi \, a_m^2 \qquad (4.1)$$

are, respectively, the scaled energy contained with the C-shaped cylinders along the array and the energy of the incident field over the same area. Therefore, $\mathcal{Q}_{\text{arr}}$ quantifies the overall amplification given by the array.

For head-on incidence, $\psi = 0$, the array amplifies the incident energy by over an order of magnitude for wavelengths $\lambda \in (60\,\text{m}, 98\,\text{m})$, with maximum amplification $\mathcal{Q}_{\text{arr}} \approx 10^{1.53} \approx 33.6$. For wavelengths $\lambda < 60$ m, the amplification reduces as wavelength decreases, due to the associated frequencies lying in quasi-bandgaps for the cylinders at the leading end of the array, so that only small quantities of wave energy penetrate the array. For wavelengths shorter than those shown, higher-order resonances alter this simple trend. For wavelengths $\lambda > 98$ m, the amplification asymptotes towards unity as the influence of the cylinder on the waves reduces. The behaviour is similar for non-head-on incidence, with some reduction in amplification, as Rayleigh–Bloch waves are not as strongly excited due to the loss of symmetry in the incident field with respect to the axis of the array. Amplifications of over an order of magnitude exist up to $\phi \approx 0.18\,\pi$, and the wavelength interval for which the amplification is over an order of magnitude is at least 25 m long up to $\phi = 0.15\,\pi$.

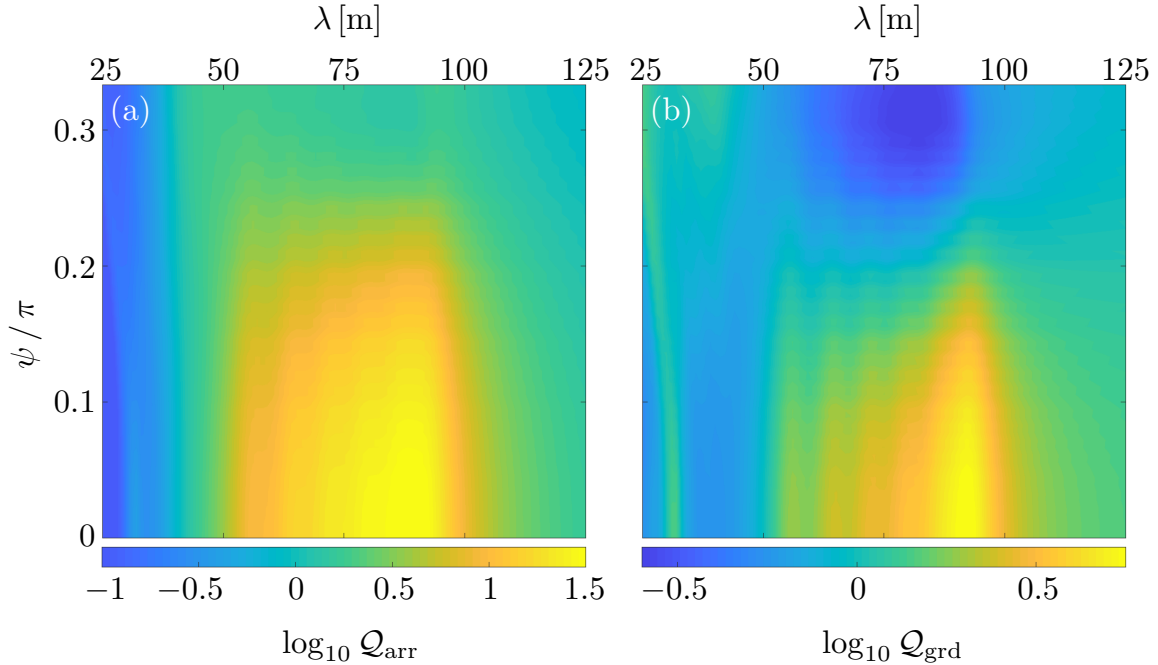

FIGURE 4. Logarithmic Q-factors as functions of incident wavelength and direction. (a) $\mathcal{Q} = \mathcal{E}/\mathcal{E}_{\mathrm{inc}}$, where $\mathcal{E}$ is the integrated energy within the C-shaped cylinders along the array, and $\mathcal{E}_{\mathrm{inc}}$ is the incident energy over the same domain. (b) $\mathcal{Q}_0 = \mathcal{E}/\mathcal{E}_0$, where $\mathcal{E}_0$ is the integrated energy within equivalent isolated C-shaped cylinders.

Fig. 4b shows $\mathcal{Q}_{\mathrm{grd}} = \mathcal{E}/\mathcal{E}_0$ on a logarithmic scale, where

$$\mathcal{E}_0 = \sum_{m=1}^{M} \iint_{\Omega_m} |\phi_m|^2 \, \mathrm{d}\mathbf{x}, \qquad (4.2)$$

with $\phi_m$ the velocity potential for the $m$th cylinder in isolation, i.e. with no surrounding cylinders. Therefore, $\mathcal{Q}_{\mathrm{grd}}$ quantifies the overall energy amplification given by the radius grading, independent of the amplification due to the cylinder resonances. As indicated by Figs. 1–2, for head-on incidence, the grading is most effective for wavelengths which excite resonances in the cylinders towards the trailing end of the array. The maximum overall amplification due to grading is $\mathcal{Q}_{\mathrm{grd}} \approx 10^{0.77} \approx 5.87$ for $\lambda \approx 92\,\mathrm{m}$. The amplification is positive for $\lambda > 50\,\mathrm{m}$, with negative amplifications associated to quasi-bandgaps for shorter waves, as noted above. Similarly, the strength of amplification due to grading slowly decreases as the incident wave direction moves away from head-on incidence, with the grading more than trebling the overall amplification, i.e. $\mathcal{Q}_{\mathrm{grd}} > 3$, for wavelengths around $\lambda = 92\,\mathrm{m}$ up to $\psi \approx 0.15\,\pi$.

## 5. Conclusions

A graded line array of C-shaped cylinders has been proposed as a structure for frequency separation and amplification of water-wave energy, and with structural di-mensions comparable to the target wavelengths. Using linear potential-flow theory, and an example in which the array consists of ten cylinders with graded radii, it was shown that the resonant amplifications within a given cylinder in the array far exceed those of the cylinder in isolation, and that typically even larger amplifications occur in the preceding cylinder (with respect to wave direction). Further, the array was shown to be effective in terms of the overall amplification, over broad ranges of wavelengths and incident directions.

A recently developed transfer-matrix solution method was employed, which provided insights into the mechanisms underlying the large amplifications. Specifically, the method was used to show the amplifications are generated by excitation of Rayleigh–Bloch waves — previously only known for arrays of regular cylinders in the water-wave context — and progressive slowing down of the Rayleigh–Bloch-wave group velocity along the array until it ceases propagating. Further, it was shown that the amplification locations can be controlled and predicted using the lowest-resonant frequencies of individual cylinders, as the resonances determine the cut-off frequencies of the Rayleigh–Bloch dispersion curves.

The authors thank Fabien Montiel for the C-shaped cylinder code used to conduct this study, and Vicente Romero-García and Jean-Philippe Groby for useful discussions. The Isaac Newton Institute for Mathematical Sciences provided support and hospitality to the authors during the programme Mathematics of Sea Ice Phenomena (EPSRC grant number EP/K032208/1), when work on this paper began. LGB was also partially supported by a grant from the Simons Foundation. RVC thanks the EPSRC and Leverhulme Trust for their support through grant EP/L024926/1 and Research Fellowship respectively.